\documentclass[12pt,english]{article}
\usepackage{graphicx}
\usepackage{epsfig}
\usepackage[english]{babel}
\usepackage[ansinew]{inputenc}
\usepackage[T1]{fontenc}
\usepackage{latexsym}
\usepackage{amssymb}

\begin{document}
\title{
Simulations of neutrino oscillations for a wide band beam from CERN to LENA}
\author{Juha Peltoniemi\\ \em Excellence Cluster Universe \\ \em Technische Universität München, Garching, Germany}
\date{\today}
\maketitle

\begin{abstract}
Neutrino oscillations from a wide band beam of 1--6 GeV at CERN to LENA, a 50 kton liquid scintillator, at Pyh\"asalmi mine 2288 km apart are simulated. The performance is very promising and this can be considered as a realistic alternative for the next long baseline experiment. Different performance factors and baselines are compared and the studied setup is found to be sufficiently close to a realistic optimum.
\end{abstract}

\section{Introduction}

LENA (Low Energy Neutrino Astronomy) has been proposed to study proton decay and low energy neutrinos from the Earth, the Sun and supernovae\cite{Wurm:2007cy, Oberauer:2006cd, Hochmuth:2006gz, MarrodanUndagoitia:2006qs, MarrodanUndagoitia:2006qn, MarrodanUndagoitia:2006rf, MarrodanUndagoitia:2006re, Undagoitia:2005uu}.  It is a large volume liquid scintillator with a nominal mass of 50 kton, at the current design a hundred metres high and 30 m wide vertical cylinder with 13 000 phototubes. 

Here is proposed to use LENA as the detector for a wide band beam from CERN.
If LENA will be built for neutrino astronomy, not using it for beam would be a waste of opportunity.

It was recently postulated \cite{Peltoniemi:2009xx} (see also \cite{Learned:2009rv}) that LENA may have a good tracking capacity and a reasonable energy resolution to be used for high energy neutrino experiments. Previously sub-GeV neutrinos from a beta beam have been simulated\cite{marrodan}. All the results are preliminary and a lot of more detailed studies --- both computational and experimental --- are required to define the performance more accurately. Moreover, the performance of LENA will depend on its design, and it is very important to make clear whether to take the beam option into account in the planning.  

A candidate site for LENA is 1444 m deep (4000 m.w.e.) Pyh\"asalmi Mine in Central Finland. The proposed location for the detector is 1450 m deep (4000 m.w.e.), and the site provides very good logistical conditions. For low-energy measurements the absense of nuclear reactors nearby is an advantage. 
The baseline from CERN to Pyh\"asalmi is 2288 km long. 

Other sites being studied in LAGUNA \cite{Autiero:2007zj} include Slanic (1544 km), Boulby (1050 km), Sierosczowicze (950 km), Canfranc (665 km), Umbria (630 km) and Frejus (130 km). Their feasibility is under study.

The density profile for the CERN-Pyh\"asalmi baseline has been modelled well \cite{Kozlovskaya:2003kk,Peltoniemi:2006hf}, and the accuracy of the average density can be taken 1 \%. The density profiles for other baselines have not been modelled, and they may be less accurate. Of these, the baseline towards Poland may be rather well modellable, but the baseline towards Romania is the most challenging, due to complicated mountain chains, and the accuracy will be worse.

\section{Beams}

\begin{figure}[tbhp]
\begin{center}
\includegraphics[angle=0, width=10cm]{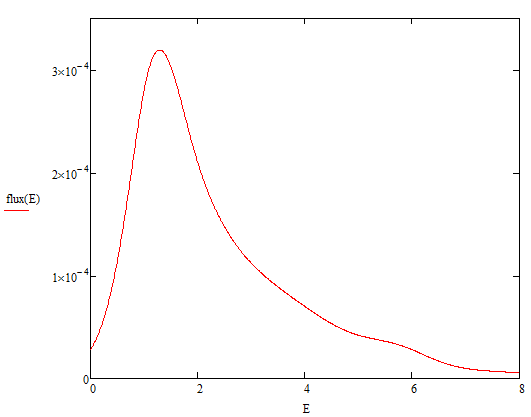}
\caption{\sf The assumed spectrum used in these simulation, in units of $\nu/$GeV m$^2$POT at 1 km distance. The beam is taken as is, not optimised for any particular distance. The electron flux is taken just 1 \% of this.
Total neutrino number is $8\cdot10^{-4} \nu/\mbox{m}^2$POT.  }
\label{nus}
\end{center}
\end{figure}

For the given distance of 2288 km the first oscillation peak is seen at about 4.2 GeV, with the usual values for neutrino mixing. To see other peaks, energies 1--3 GeV are desired.

The most cost-effective beam would be a conventional wide band beam of 1--6 GeV. Such a beam could be produced by SPS (400 GeV), similarly as the CNGS beam but tuned for lower energies. Another alternative would be PS2 (50 GeV). The cost of the beam, no more than O(100 MEUR), is marginal compared with a beta beam or a neutrino factory. 
	
In this work I assume the neutrino spectrum as depicted in figure \ref{nus}, adapted from simulations from Fermilab or BNL\cite{BNL,Fermilab} (also in line with some simulations for CNGS beam \cite{Rubbia:2002rb, Meregaglia:2006du, Ferrari:2002yj}). This is just an ad hoc assumption, not optimised for anything. I take the total neutrino number to be $8\cdot10^{-4} \nu/\mbox{m}^2$ POT, and the beam power $3.3\cdot 10^{20}$ POT/a or 1.5 MW. Running time is $5+5$ years. The beam is the least accurately known experimental feature, and factor of 2 differences may appear in the shape of the spectrum, depending on the design of the pion focusing system, and the beam power is subject to future decisions.

I consider only on-axis beam. While off-axis might give a slightly better energy spectrum, on the downside there will be more beam induced electron neutrino background. Further simulations should show whether a very small off-axis angle would be a viable compromise.

\section{Oscillations}

\begin{figure}[tbhp]
\begin{center}
\includegraphics[angle=0, width=10cm]{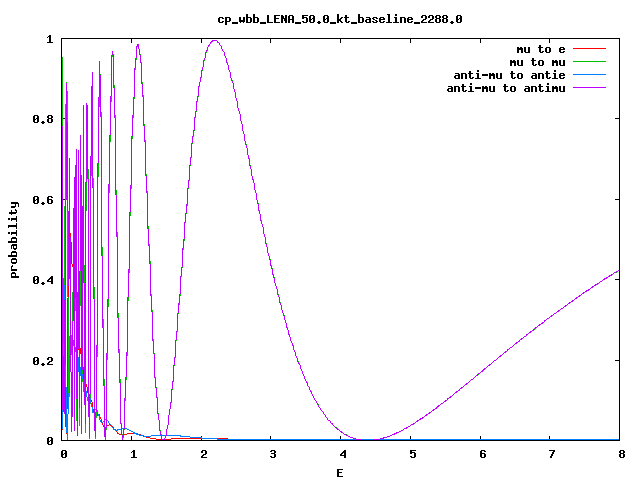}
\caption{\sf Sample plots of simulated oscillations with the 2288 km long beamline, for $\delta = \pi/2$ and $\sin^2 2\theta_{13}= 10^{-2.5}$. The first minimum is at $4.2 \pm 0.3$ GeV. Evidently a wide band beam 2-5 GeV would be optimum for this baseline.}
\label{oscillations}
\end{center}
\end{figure}

I assume for the oscillation parameters the following values \cite{Amsler:2008zzb}:
\begin{eqnarray}
\Delta m_{21}^2&=& (7.59 \pm 0.21) \cdot 10^{-5} \mbox{ eV}^2 \\ 
\Delta m_{31}^2&=& (2.43 \pm 0.13) \cdot 10^{-3} \mbox{ eV}^2 \\
\sin^2 2\theta_{13} &<& 0.074\\
\sin^2 2\theta_{12} &=& 0.87 \\
 \theta_{12} &=& 0.60 \pm 0.02  \\ 
\sin^2 2\theta_{23} &=& 0.99 \\
\theta_{23} &=& \pi/4 \pm 0.06, 
\end{eqnarray}
the last to cover all octants. 
The errors to be used might be smaller than these: what we should use in simulations is the expected uncertainty at the time of analysis and not our present igrorance, but I cannot predict that any better. 

The signals in a wide band beam experiment are electron neutrino appearance and muon neutrino disapperance. Evidently the first one gives stronger contribution. 

It is known (see e.g. \cite{Barger:2006vy} ) that baselines 1200--2500 km are optimal for a few GeV wide band beam. Particularly, a baseline 2000--2500 km is best for mass hierarchy, while for CP-violation searches shorter baselines 1300 km have been considered to be slightly better (with fixed beam, though). To resolve the $\theta_{23}$ octant, baselines longer than 2000 km are required. For theta reach the optimum is between 1200--2300 depending on the CP angle. These prejudices of course are very sensitive to all assumptions on beam composition and detector mass and resolution.

For simulating the neutrino oscillation, the standard GLoBES toolback \cite{Huber:2007ji,Huber:2004ka} is used, embedded within an own code. The density accuracy is assumed to be 1\%, unless otherwise stated. (Using other values for the density error showed no visible difference for any plot.)

As usual:
\begin{itemize}
\item The $\theta$-reach was studied comparing non-zero "true values" to "test values" with $\theta=0$, varying $\delta$ and marginalizing over all other parameters, taken into account different hierarchies.
\item Mass hierarchy: I compared $\delta m_L>0$ and $-\delta m_L + \delta m_S <0$, marginalising over all parameters. 
\item Range to observe CP-violation: I set $\delta$ to non-zero and compare with $\delta=0$ and and $\delta=\pi$, marginalising over other parameters, with both hierarchies. 
\end{itemize}

The plots are given for $\chi$. The typical corresponence is 
$3\sigma \to \chi < 9$, 
$2\sigma \to \chi < 4$ and
$1\sigma \to \chi < 1$, 
as recommended in \cite{Bandyopadhyay:2007kx}.

\section{Detector and performance}

\subsection{Description}

LENA has been proposed for low energy neutrino astronomy and proton decay. It is a 100 m high vertical detector, which is currently planned to consist of 
\begin{enumerate}
	\item 50 kton of liquid scintillator, in a tank with a radius of 12 m. Several liquids with different properties are being considered (e.g. PXE, LAB). 
	\item 20 kton of buffer in 2 m wide layer around the scintillating oil.  The buffer is similar oil but without scintillating component. However, the light yield for high-energy particles is non-zero. Buffer and scintillator are separated with a nylon vessel, and the buffer is contained in a steel vessel.
	\item 100 kton of water around the tank, to shield from neutrons.
\end{enumerate}
These dimensions may still be changed a little, depending on the chosen liquids and the tank structure.
There will be some 13 000 phototubes facing inwards, and about 1500 facing outwards to the shielding liquid. These numbers may be changed if smaller phototubes are used. The photocoverage might also be increased from 30 \% up to 70 \% if increasing the number of phototubes and using light collectors.

The fiducial mass of LENA is planned to be 50 kton. The fiducial mass for high-energy neutrinos may differ from that for low-energy neutrinos. Particularly, the fiducial volume decreases with longer tracks, and I assume the efficiency to go linearly to zero from 3 GeV to 7 GeV for muons.

It may be possible to use the external water shielding for additional volume to track the highest-energy particles by Cherenkov light. Water might even be replaced by cheap scintillation oil. Also the 2 m buffer will produce scintillation and Cherenlov light, albeit much less than the proper scintillator. The resolutions of the external buffer and shield may, however, be substantially poorer than that of the internal fiducial volume, depending on the instrumentation.

To study the dependence on the statistics I made several runs using the detector mass as a scaling variable. It should be understood that in those studies the varying fiducial mass may include also variations in beam power, running time and detection efficiency. For comparison I also study the case of horizontal layout, aligned with the beam. This is implemented by different efficiency for high-energy muons.

\subsection{Performance}

The performance of LENA for measuring high-energy neutrinos is being studied, and some intermediate results were released recently \cite{Peltoniemi:2009xx}. It was found that LENA has a good tracking capacity, with an ability to distinguish and measure at least three tracks, if they are sufficiently long and well separated. To achieve this, however, we need a photodetection system with good multiphoton capacity and a fast scintillator.

The particle identification is very good. Lepton flavor can be defined reliably, and there is a limited capacity to distinguish neutrinos from antineutrinos using different signatures of protons and neutrons, though only statistically due to related nuclear physics. 
The spatial and angular resolutions are also good, although for beam physics they have only indirect relevance.

The energy resolution is rather complicated. The light output can be measured in theory with accuracy 0.3--0.1 \%, and allowing some desperfections or saturation in photosensors maybe still 1 \%. Hence in practice most errors are due to deviations from linear response between energy loss and light emission. Other than the instrumental effects, the deviations may be due to:
\begin{enumerate}
	\item Light attenuation within the scintillator --- depends on the position of the emission. A good positional resolution is needed.
	\item Quenching i.e. reduced light yield for a large local energy deposit. The quenching is largest for non-relativistic particles, like protons and alphas. Good particle identification is necessary, as well as reliable recognition of all particles, including low-energy secondary particles.
	\item Nuclear physics --- partly unavoidable.
\end{enumerate}	
Good understanding on the physics of the neutrino collision may improve the energy resolution, e.g. by taking into account correlations between incident neutrino energy and the energies and scattering angles of the scattered particles.
I estimate for the errors in energy measurement:
\begin{enumerate}	  
	\item Accuracy of the tracking analysis: assumed 1 \% throughout the range.
	\item Nuclear physics related to the scattering from a nucleon bound in carbon nucleus. This is not easily estimated, and as guidelines we may use the binding energy of last nucleon, 16.0 or 18.7 MeV, or Fermi energy of Fermi bag model, 37 MeV. Here I take the ($1\sigma$) uncertainty to be 20 MeV. 
	\item Other stochastic effects not taken into account, particularly related to hadrons and nuclear fractions, including neutrons. I take for it 10--50 MeV$\sqrt{E}$, depending on case.
\end{enumerate}
Hence I assume, purpotedly optimistically, for the energy resolution for the different event categories:
\begin{eqnarray}\label{eres}
\delta E_{\rm ep} &=& 0.01 E + 30 \mbox{ MeV} \sqrt{\frac{E}{1 \mbox{ GeV}}}+ 20 \mbox{ MeV}\\
\delta E_{\rm en} &=& 0.01 E + 50 \mbox{ MeV} \sqrt{\frac{E}{1 \mbox{ GeV}}}+ 20 \mbox{ MeV} \\
\delta E_{\rm \mu p} &=& 0.01 E + 10 \mbox{ MeV} \sqrt{\frac{E}{1 \mbox{ GeV}}}+ 20 \mbox{ MeV}\\
\delta E_{\rm \mu n} &=& 0.01 E + 40 \mbox{ MeV} \sqrt{\frac{E}{1 \mbox{ GeV}}}+ 20 \mbox{ MeV}
\end{eqnarray}
Nevertheless, to study the dependence on the resolution, I repeat the studies with varying energy resolution, using a simple linear formula $\delta E = \alpha E$ with $\alpha$ a free (continuous) parameter. 

For the background, I assume:
\begin{itemize}
	\item Electron neutrino appearance channel: The 1 \% beam contamination is dominant and completely unavoidable with any detector technology. Compared with this, the misidentification of lepton flavor is negligible. Neutral current background ($\pi^0 \to \gamma\gamma$) is partly identifiable and probably less than the beam background. However, the neutral current background depends a lot on the beam, particularly on its high-energy tail. As this is not known precisely, I just use the same values as for NOvA\cite{Ambats:2004js,Yang_2004}, i.e. 0.15 -- 0.37 \%. 
	\item Muon neutrino disappearance channel: The neutrino beam has 2 \% contamination of muon antineutrinos (or vice versa). By efficient particle reconstruction procedure the beam background may be reduced which is studied as an optional feature. Flavor misidentification is negligible, and also the charged pion production by neutral currents is probably less relevant though not completely ignorable.
\end{itemize}

The efficiency for muon neutrinos is assumed 100 \% and for electron neutrinos 90 \%. A smaller overall efficiency would count like decreasing the fiducial mass of the detector.

\section{Results}

\subsection{CP Violation}

\begin{figure}[tbhp]
\begin{center}
\includegraphics[angle=0, width=10cm]{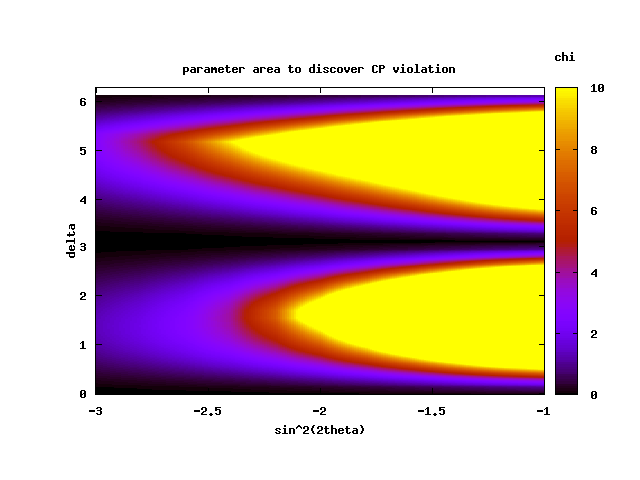}
\caption{\sf The discovery range of the CP angle, with 50 kton LENA with standard performance at 2288 km baseline. The color maps the $\chi$-values, so that the yellow areas are about $3\sigma$, red $2\sigma$ and blue $1\sigma$.}
\label{deltas}
\end{center}
\end{figure}

Searching for the CP violation is the most important task of the long baseline study. As seen in Fig.~\ref{deltas},
with the default parameters the discovery potential of a non-zero CP angle extends to $\sin^2 2 \theta_{13}>3\cdot 10^{-3}$. The fraction of CP angles coverable for $\sin^2 2 \theta_{13} = 0.1$ is 67.5 \%. 

\begin{figure}[tbhp]
\begin{center}
\includegraphics[angle=0, width=7cm]{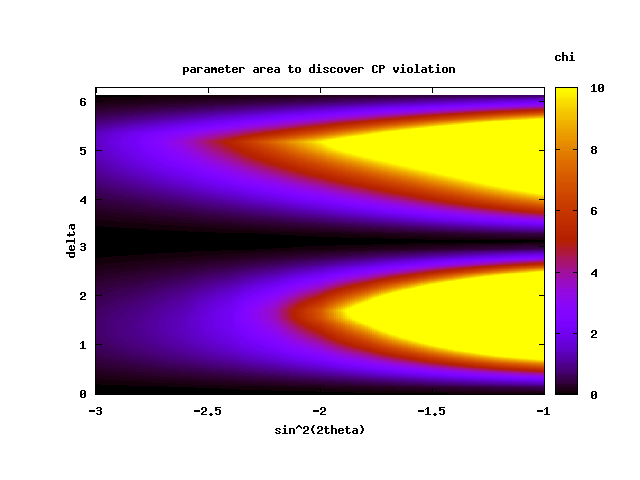}
\includegraphics[angle=0, width=7cm]{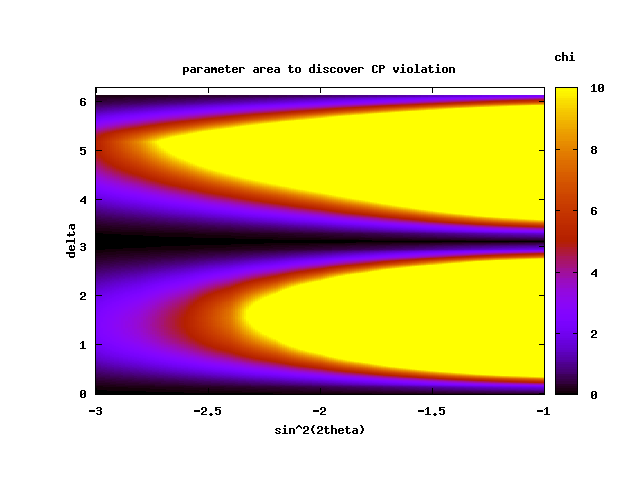}
\caption{\sf Comparison of CP range for 25 kton, and 100 kton fiducial masses, with 2288 km baseline. The differences are significant.}
\label{deltasdet}
\end{center}
\end{figure}

The size of the detector (or beam power) is important, with 25 kton the performance is much worse (Fig.~\ref{deltasdet}), and larger sizes --- at least up to 300 kton --- improve the capacity significantly but not drastically. 

\begin{figure}[tbhp]
\begin{center}
\includegraphics[angle=0, width=7cm]{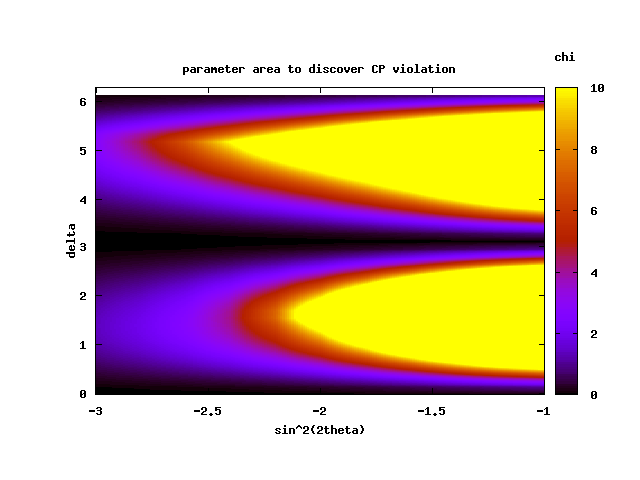}
\includegraphics[angle=0, width=7cm]{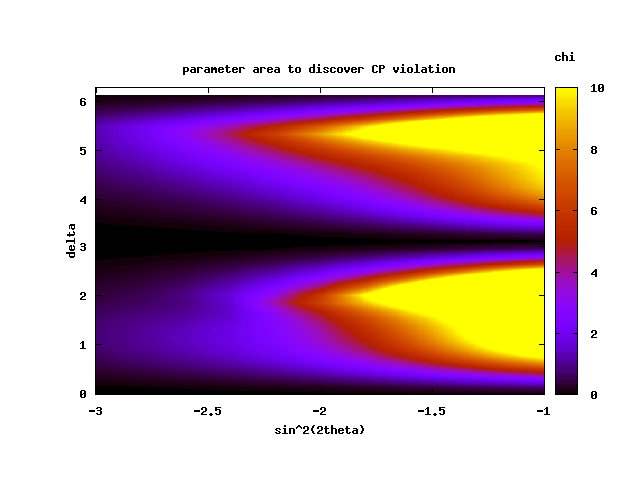}
\caption{\sf Comparison of CP range with 5 \%, and 20 \% relative energy resolutions. The results are not sensitive to the energy resolution better than 5 \%, but worsening the resolution beyond 10 \% weakens the performance.}
\label{deltasdetq}
\end{center}
\end{figure}

An adequate energy resolution is very important (Fig. \ref{deltasdetq}), a resolution worse than 10 \% spoils the performance. However, improving the resolution better than five per cent does not give any significant benefit.

I made some comparisons with different baselines. It was found that baselines 1200--3000 km are quite equal in performance, but shorter baselines suffer from the poor determination of the mass hierarchy. A set of two baselines --- longer for mass hierarchy and shorter to scrutinize CP violation --- may be more optimal though not that realistic.


\subsection{Mass hierarchy}

\begin{figure}[tbhp]
\begin{center}
\includegraphics[angle=0, width=10cm]{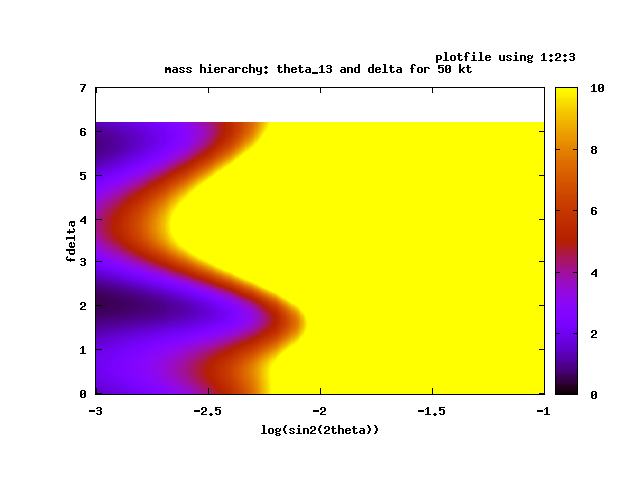}
\caption{\sf The capacity to measure the mass hierarchy with  different values of the mixing angle $\theta_{13}$ and CP angle $\delta$ for a 50 kton LENA (LS).}
\label{hierarchy}
\end{center}
\end{figure}

\begin{figure}[tbhp]
\begin{center}
\includegraphics[angle=0, width=10cm]{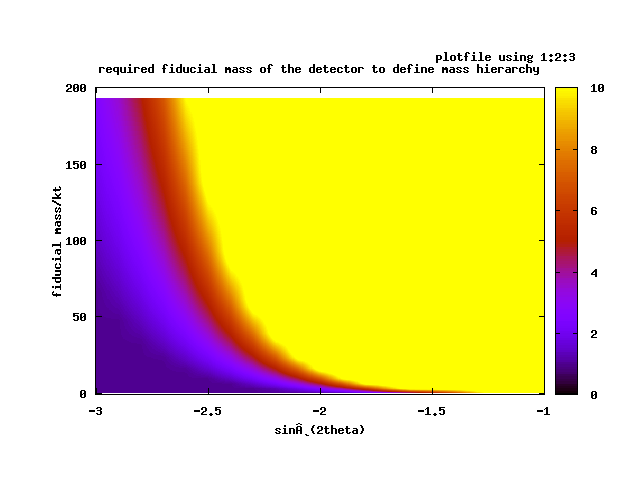}
\caption{\sf The capability to measure the mass hierarchy for different fiducial masses of the detector.  The fiducial mass is used as a scaling variable, absorbing also variations in detection efficiency and beam power. No other changes, e.g. in efficiency profile are assumed.}
\label{mhperf}
\end{center}
\end{figure}

\begin{figure}[tphb]
\begin{center}
\includegraphics[angle=0, width=10cm]{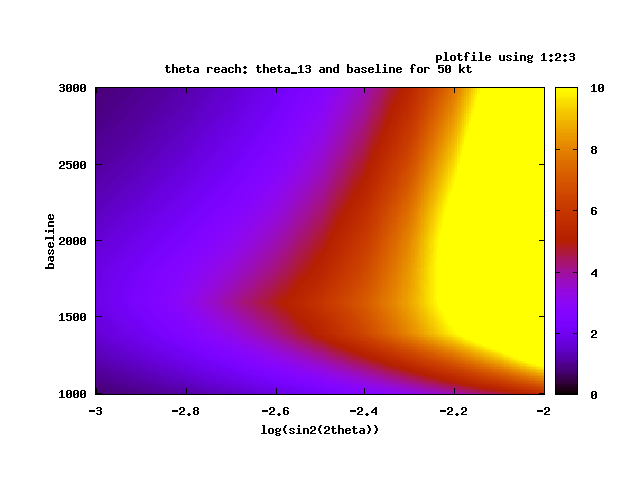}
\caption{\sf Comparison of the performance of the mass hierarchy for different baselines. Here 1600 km looks the optimum, though baselines 1400--2500 km are quite equal. The differences may be partly due to assumed ad hoc beam spectrum that peaks at 1.5 GeV, favouring lengths just below 1000 km.}
\label{mhbaselines}
\end{center}
\end{figure}

\begin{figure}[tbhp]
\begin{center}
\includegraphics[angle=0, width=10cm]{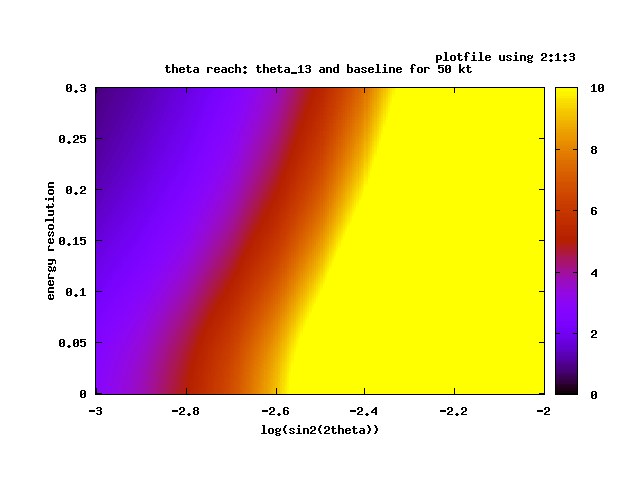}
\caption{\sf Comparison of the performance of the mass hierarchy for different relative energy resolutions (0--30 \%). The relation is rather simple. Some 5 \% is clearly sufficient and additional improvements will not improve the performance significantly.}
\label{mheres}
\end{center}
\end{figure}

The nextmost important (except possible surprises) is to define the sign of  $\delta m_{23}$. 
The results are plotted in the figures \ref{hierarchy}--\ref{mheres}. 
For a 50 kton detector at 2288 km one can define the mass hierarchy up to angles 
$ \sin^2 (2 \theta) > 10^{-2}$ (for any $\delta$), or $ \sin^2 (2 \theta) > 2\cdot10^{-3}$ for the most optimal $\delta$.

The detector size 50--100 kton is very good: Smaller detectors are limited by statistics and larger by systematics. It is hard to go beyond 
$ \sin^2 (2 \theta) \sim 2\cdot10^{-3}$ with any realistic size of the detector.
At small $\theta_{23}$ the beam background is the most constraining factor.

When comparing different baselines (Fig.~\ref{mhbaselines}), lengths 1200--3000 km perform quite similarly, but baselines shorter than 1000 km are significantly worse. The optimal was 1600 km, but the small differences may well be due to the assumed beam spectrum that was not optimised for anything.

The measurement of the mass hierarchy is rather weakly sensitive to the detector quality and orientation. The rejection of wrong sign muon background plays little role, and improving the energy resolution better than 5 \% is irrelevant (Fig.~\ref{mheres}).

\clearpage

\subsection{Mixing angles}

\begin{figure}[tbp]
\begin{center}
\includegraphics[angle=0, width=10cm]{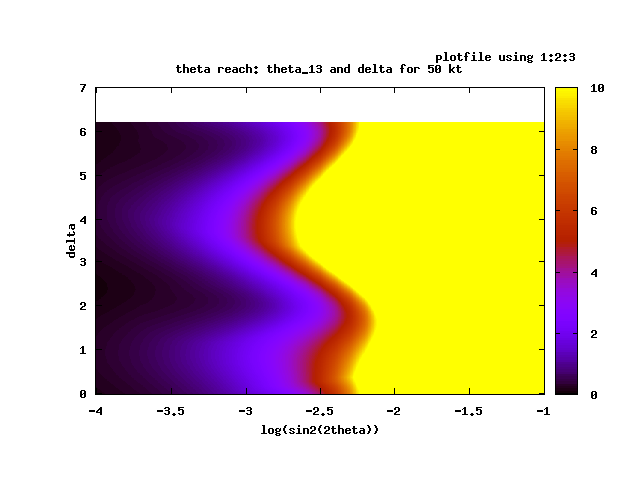}
\caption{\sf The $\theta_{13}$ reach ($3\sigma$) in $\delta$-plot for 50 kton LENA at 2288 km baseline. }
\label{reach}
\end{center}
\end{figure}

\begin{figure}[hbtp]
\begin{center}
\includegraphics[angle=0, width=10cm]{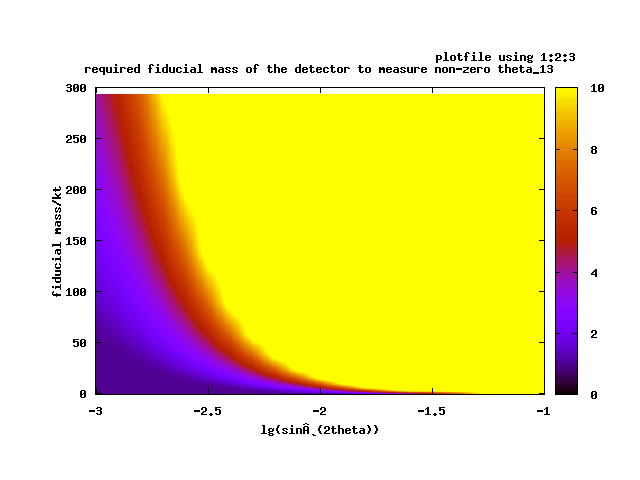}
\caption{\sf The $\theta_{13}$ reach ($3\sigma$) as a function of the fiducial mass of the detector at 2288 km baseline.}
\label{reachfid}
\end{center}
\end{figure}

With a 50 kton detector at Pyh\"asalmi mine, the range to measure $\theta_{13}$ extends to mixing angles up to $\sin^2 2\theta_{13} > 6\cdot 10^{-3}$ for any $\delta$. With the optimal delta one may reach $\sin^2 2\theta_{13} > 2\cdot 10^{-3}$.

When comparing the performance with the fiducial mass of the detector, we observe that there is a steep sharp rise after $ \sin^2 (2 \theta) > 3\cdot 10^{-3}$ and one cannot go much beyond that with any realistic detector using a realistic wide band beam. Reaching $\sin^2 (2 \theta) \sim 10^{-3}$ would require a 500 kton detector or an order-of-magnitude stronger beam.

\begin{figure}[htbp]
\begin{center}
\includegraphics[angle=0, width=7cm]{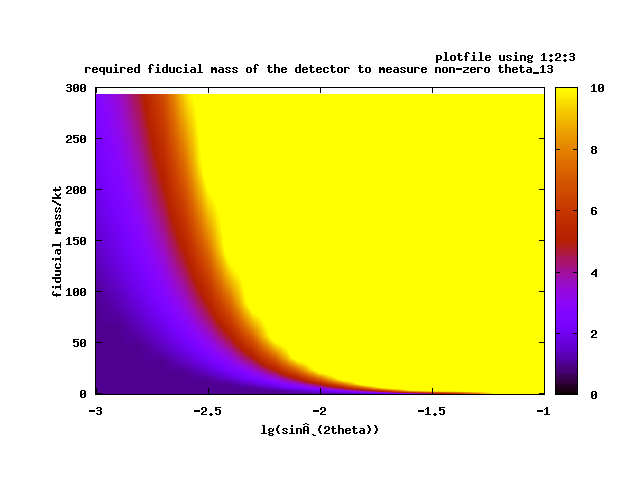}
\includegraphics[angle=0, width=7cm]{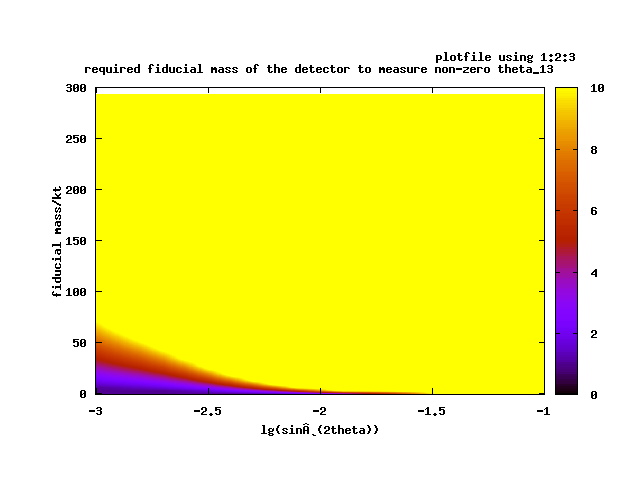}
\caption{\sf Studies about different backgrounds. In the left a run with neutral current background set ten times larger than the default. A run without any neutral current background is not shown because it is almost equal to the default case. In the right side is a plot from a run with no background at all, not even beam background. While completely unphysical, this reveals that the beam background is the bottleneck for small angles.}
\label{reachNB1}
\end{center}
\end{figure}

Improved detector resolution or background rejection do not help much. A run without background (Fig. \ref{reachNB1}) reveals the bottleneck: the smaller angle performance is limited by the beam background. To study small angles one needs a cleaner beam, like a high-intensity beta beam or a neutrino factory.

The capacity to define the octant appears very limited. This must be studied more. Because of this no octant was assumed for $\theta_{23}$ but it was allowed to vary over $\pi/4$.

\begin{figure}[tbhp]
\begin{center}
\includegraphics[angle=0, width=10cm]{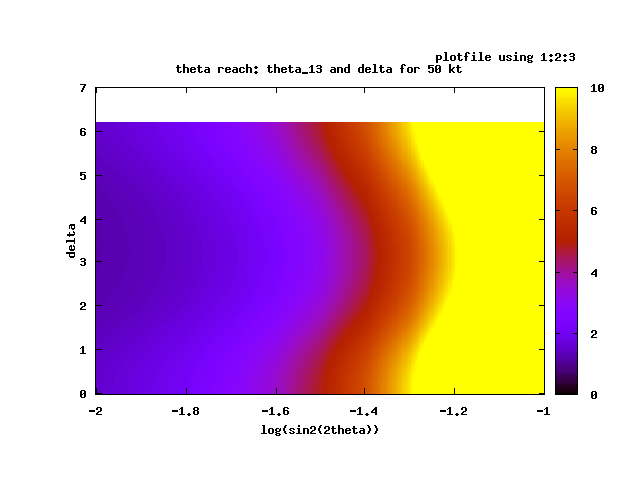}
\caption{\sf The range to define the octant of $\theta_{23}$ for 50 kton LENA (2288 km).}
\label{octant}
\end{center}
\end{figure}

\begin{figure}[tbhp]
\begin{center}
\includegraphics[angle=0, width=10cm]{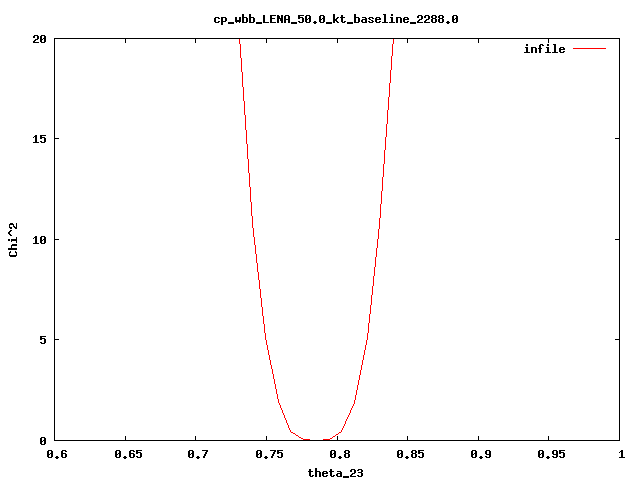}
\caption{\sf The accuracy of $\theta_{23}$ for 50 kton LENA, with $\sin^2 2\theta_{13} > 3\cdot 10^{-3}$.}
\label{acc23}
\end{center}
\end{figure}

This long baseline experiment can significantly improve the accuracy of $\theta_{23}$, as shown in Fig.~\ref{acc23}.

\clearpage

\subsection{Case of large $\theta_{13}$}

By the time of this experiment running we should know the results of Double Chooz cite{Ardellier:2006mn}. 
The sensitivity of Double Chooz is about $\sin^2(2\theta_{13}) \sim 0.02$. 
If Double Chooz measures $\theta_{13}$, we have very good potential to discover the CP violation. The mass hierarchy will be discovered unambiguously for even the most pessimistic detector setup.

\begin{figure}[tpb]
\begin{center}
\includegraphics[angle=0, width=7cm]{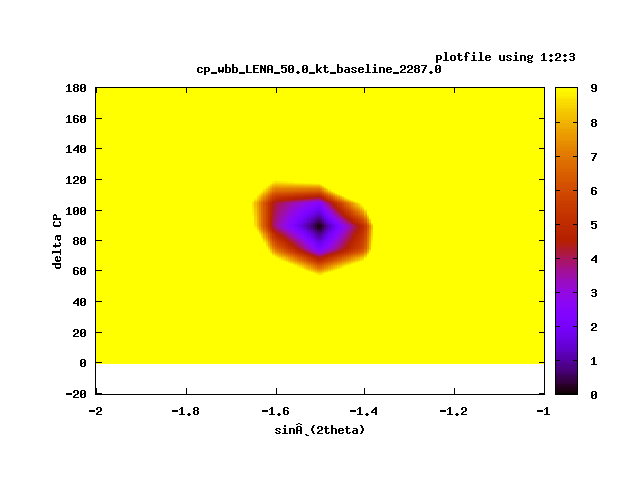}
\includegraphics[angle=0, width=7cm]{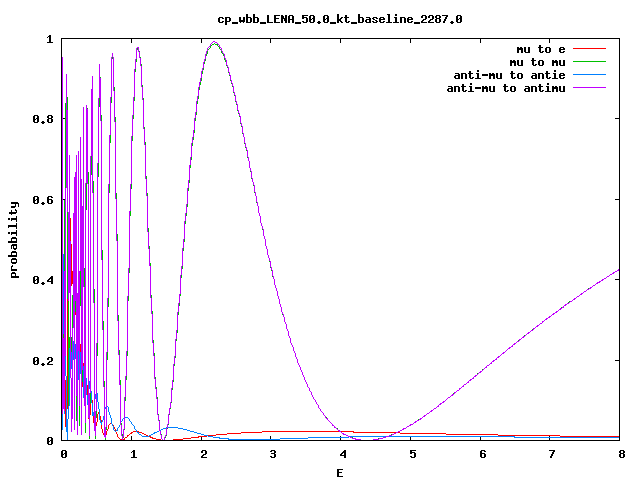}
\includegraphics[angle=0, width=7cm]{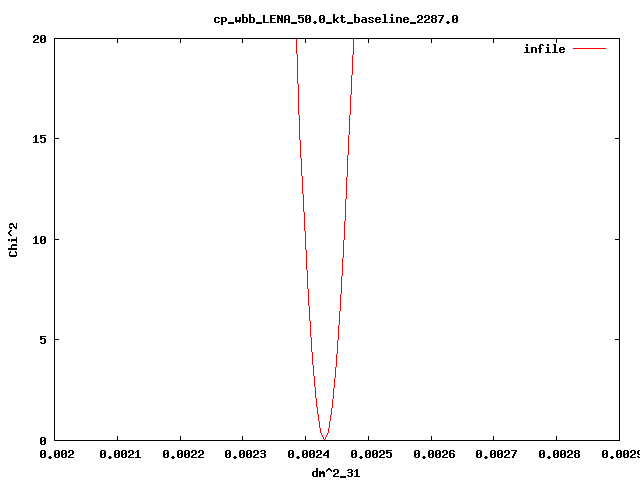}
\includegraphics[angle=0, width=7cm]{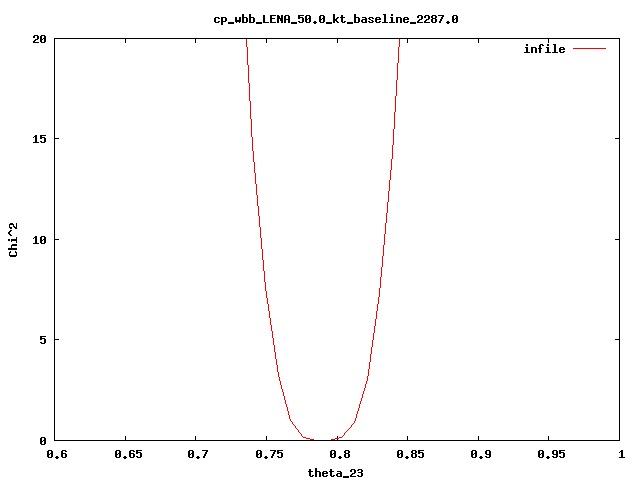}
\caption{\sf Some plots for the case $\sin^2(2\theta_{13}) \sim 0.03$, at 2288 km. First row left: Accuracy of $\theta_{13}$ and $\delta$. Right: Oscillation curves. Second row left: Accuracy of $\delta m^2_{23}$. Right: accuracy of $\theta_{23}$. The width of the plots is approximately the current $3\sigma$ limit.}
\label{large}
\end{center}
\end{figure}

More exactly, in the case of 
$\sin^2(2\theta_{13}) \sim 0.03$ (Fig. \ref{large}):
\begin{itemize}
\item The mass difference $\Delta m^2_{31}$ will be measured at precision of $0.04\cdot10^{-3} \mbox{eV}^2 (3\sigma)$.
\item The angle $\theta_{13}$ will be determined at good accuracy, $\pm 0.2$ for $\sin^2(2\theta_{13})$.
\item The angle $\theta_{23}$ will be measured with good accuracy, $\pm 0.05 (3\sigma)$.
\end{itemize}

\clearpage

\section{Conclusions}

The results hint that LENA may be a very good option for a far detector of a wide band beam. Although the detector performance is not yet known very accurately, it is evidently sufficient even with the most pessimistic assumptions, if the beam option is approriately taken into account in its design. The results are obvisously generalisable to other detectors with similar properties.

It is evident that the detector mass 50--100 kton is very good for this beam. Smaller detectors are statistically limited, larger systematically, by the beam background. Increasing the beam power or the detector size would be desired if the discovery of $\theta_{13}$ or $\delta$ is just around the corner. If $\theta_{13}$ is too small to see a trace of it, however, we would need another, cleaner beam (and only in that case). 

For the wide band beam the vertical orientation is not a significant burden, because the main signal is electron neutrinos with 2--3 m shower length. The vertical attitude is similar to removing a couple of ktons of fiducial mass (probably like 5--10 kton --- further geometrical simulations necessary to quantify) and it is probably not worth the additional effort and increased cost to rerotate it horizontal. Moreover, it may be possible to regain substantial additional fiducial volume (up to 120 kton) by using the buffer and the external shield as additional active volumes.

The energy resolution of LENA ---  even with the most pessimistic assumptions --- is sufficient. Better resolution would not increase the performance significantly. Poorer resolution (beyond 10 \%), however,  decreases the capacity, particularly for discovering the CP violation.

The identification capacity of LENA is evidently sufficient to eliminate most of the detector-dependent background. As the dominant background is the beam contamination, the requirements for the detector are not excessive. The background due to lepton misidentification is ignorable.
The neutral current background is important if it is above 1 \% level. LENA may have some capacity to distinguish neutral current events (high-energy pions faking leptons), but the overall background level depends substantially on the beam spectrum which is not defined too well. While LENA may have some capacity to statistically discriminate neutrinos and antineutrinos, it has little relevance for the success of the expected discoveries (but that may be important for successive precision measurements). 

The most unknown component of this study is the beam. The beam power and quality depend on the design of the horns and decay line as well as the decisions to upgrade the proton chain of CERN. Both upgraded PS and SPS would do. 

The 2288 km long baseline is very powerful for studying the mass hierarchy and the CP violation. Other baselines longer than 1000 km will also perform fine, but the true optimum may depend on the beam spectrum and the detector performance.

We may conclude that LENA is an optimal detector for wide band beams. A wide band beam from CERN to LENA at Pyh\"asalmi Mine is a logical and cost-effective choice for the next long baseline neutrino oscillation experiment, whatever is the result of Double Chooz.
\\ 

{\bf Acknowledgement}\\
This research was supported by the DFG cluster of excellence "Origin and Structure of the Universe". I thank LENA collaboration, LAGUNA consortium and GLoBES team (particularly Joachim Kopp) for inspiring co-operation, advice and help and E15 group of Franz von Feilitzsch for hospitality.



\begin{thebibliography}{10}

\bibitem{Wurm:2007cy}
M. Wurm et~al.,
\newblock Phys. Rev. D75 (2007) 023007, astro-ph/0701305.

\bibitem{Oberauer:2006cd}
L. Oberauer et~al.,
\newblock Prepared for 3rd International Workshop on NO-VE: Neutrino
  Oscillations in Venice: 50 Years after the Neutrino Experimental Discovery,
  Venice, Italy, 7-10 Feb 2006.

\bibitem{Hochmuth:2006gz}
K.A. Hochmuth et~al.,
\newblock Earth Moon Planets 99 (2006) 253, hep-ph/0610048.

\bibitem{MarrodanUndagoitia:2006qs}
T. Marrodan~Undagoitia et~al.,
\newblock J. Phys. Conf. Ser. 39 (2006) 287.

\bibitem{MarrodanUndagoitia:2006qn}
T. Marrodan~Undagoitia et~al.,
\newblock J. Phys. Conf. Ser. 39 (2006) 269.

\bibitem{MarrodanUndagoitia:2006rf}
T. Marrodan~Undagoitia et~al.,
\newblock Prog. Part. Nucl. Phys. 57 (2006) 290.

\bibitem{MarrodanUndagoitia:2006re}
T. Marrodan~Undagoitia et~al.,
\newblock Prog. Part. Nucl. Phys. 57 (2006) 283, hep-ph/0605229.

\bibitem{Undagoitia:2005uu}
T. Marrodan~Undagoitia et~al.,
\newblock Phys. Rev. D72 (2005) 075014, hep-ph/0511230.

\bibitem{Peltoniemi:2009xx}
J. Peltoniemi,
\newblock (2009), 0909.4974.

\bibitem{Learned:2009rv}
J.G. Learned,
\newblock (2009), 0902.4009.

\bibitem{marrodan}
T. Marrodan~Undagoitia,
\newblock {PhD thesis}, {Technische Universit\"at M\"unchen}, 2008.

\bibitem{Autiero:2007zj}
D. Autiero et~al.,
\newblock JCAP 0711 (2007) 011, 0705.0116.

\bibitem{Kozlovskaya:2003kk}
E. Kozlovskaya, J. Peltoniemi and J. Sarkamo,
\newblock (2003), hep-ph/0305042.

\bibitem{Peltoniemi:2006hf}
J.T. Peltoniemi and J. Sarkamo,
\newblock Nucl. Phys. Proc. Suppl. 155 (2006) 201.

\bibitem{BNL}
BNL neutrino group.

\bibitem{Fermilab}
M. Bishai et~al.,
\newblock Fermilab preprint BNL-76997-2006-IR (2006).

\bibitem{Rubbia:2002rb}
A. Rubbia and P. Sala,
\newblock JHEP 09 (2002) 004, hep-ph/0207084.

\bibitem{Meregaglia:2006du}
A. Meregaglia and A. Rubbia,
\newblock JHEP 11 (2006) 032, hep-ph/0609106.

\bibitem{Ferrari:2002yj}
A. Ferrari et~al.,
\newblock New J. Phys. 4 (2002) 88, hep-ph/0208047.

\bibitem{Amsler:2008zzb}
Particle Data Group, C. Amsler et~al.,
\newblock Phys. Lett. B667 (2008) 1.

\bibitem{Barger:2006vy}
V. Barger et~al.,
\newblock Phys. Rev. D74 (2006) 073004, hep-ph/0607177.

\bibitem{Huber:2007ji}
P. Huber et~al.,
\newblock Comput. Phys. Commun. 177 (2007) 432, hep-ph/0701187.

\bibitem{Huber:2004ka}
P. Huber, M. Lindner and W. Winter,
\newblock Comput. Phys. Commun. 167 (2005) 195, hep-ph/0407333.

\bibitem{Bandyopadhyay:2007kx}
ISS Physics Working Group, A. Bandyopadhyay et~al.,
\newblock (2007), 0710.4947.

\bibitem{Ambats:2004js}
NOvA, I. Ambats et~al.,
\newblock (2004), hep-ex/0503053.

\bibitem{Yang_2004}
NOvA, T. Yang and S. Woijcicki,
\newblock (2004), Off-Axis-Note-SIM-30.

\end{thebibliography}

\appendix
\section{Additional figures and comparisons}

\subsection{Fiducial mass of the detector}

\begin{figure}[tbhp]
\begin{center}
\includegraphics[angle=0, width=7cm]{CPrange25.png}
\includegraphics[angle=0, width=7cm]{CPrange0.png}
\includegraphics[angle=0, width=7cm]{CPrange100.png}
\includegraphics[angle=0, width=7cm]{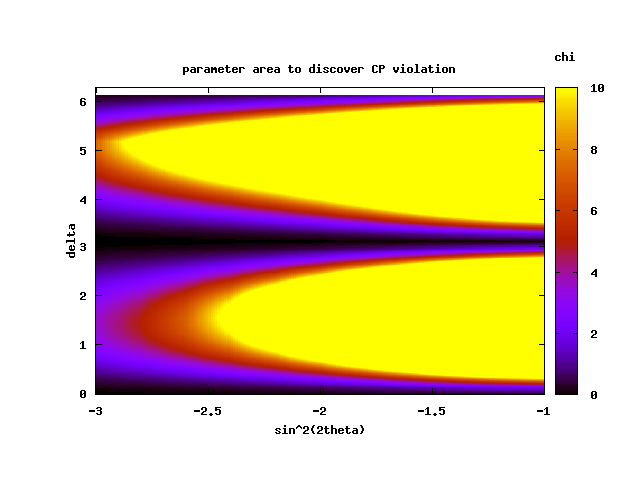}
\includegraphics[angle=0, width=7cm]{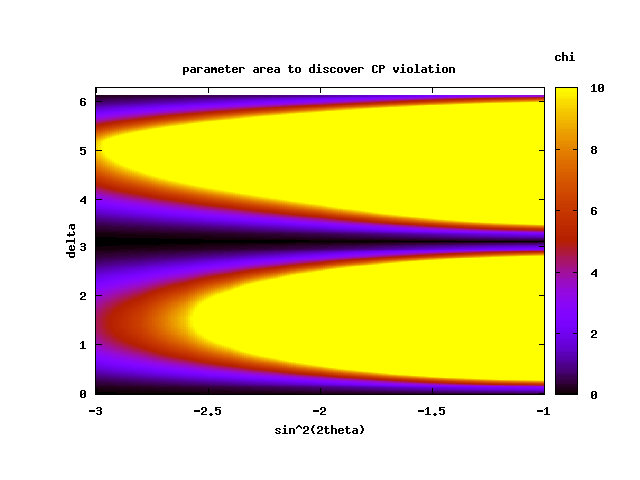}
\includegraphics[angle=0, width=7cm]{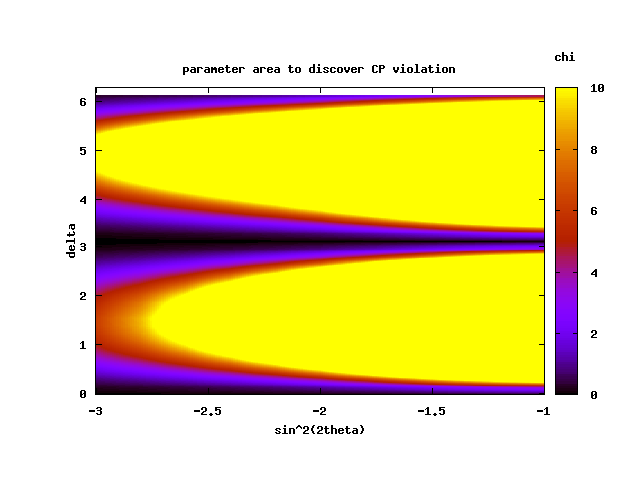}
\caption{\sf Comparison of the capacity to measure CP violation range for 25 kton, 50 kton and 100 kton, 150 kton, 200 kton and 300 kton fiducial masses, with 2288 km baseline. The differences are significant.}
\label{deltasdet2}
\end{center}
\end{figure}

\clearpage
\subsection{Horizontal vs vertical alignment}

For the vertical orientation it is assumed that the efficiency drops down at higher energies. Typically this is equivalent to reducing the fiducial volume for energies higher than 3 GeV, down to zero at 7 GeV. For electron tracks the fiducial volume is decreased by 5--10 kton. The relative cut would be smaller if the buffer or shield could be used as additional fiducial volume. That option has to be studied in more detail. 

\begin{figure}[hbp]
\begin{center}
\includegraphics[angle=0, width=7cm]{theta_reach_plot.png}
\includegraphics[angle=0, width=7cm]{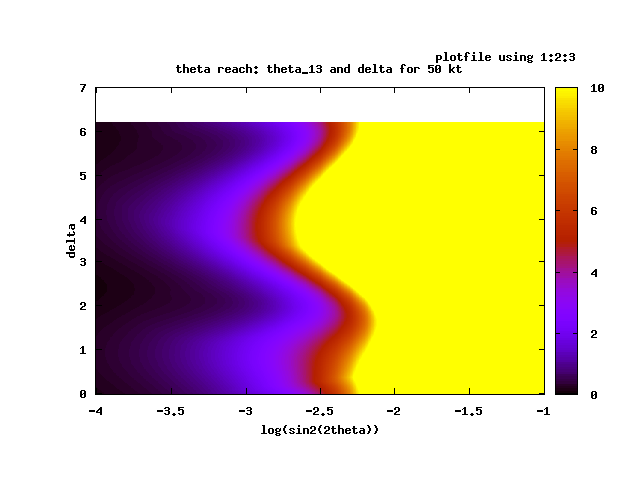}
\caption{\sf The $\theta_{13}$ reach ($3\sigma$) in $\delta$-plot for 50 kton LENA at 2288 km baseline, for vertical and horizontal orientations.}
\label{reachVH}
\end{center}
\end{figure}

\begin{figure}[tbhp]
\begin{center}
\includegraphics[angle=0, width=7cm]{theta_reach_fid_L0.png}
\includegraphics[angle=0, width=7cm]{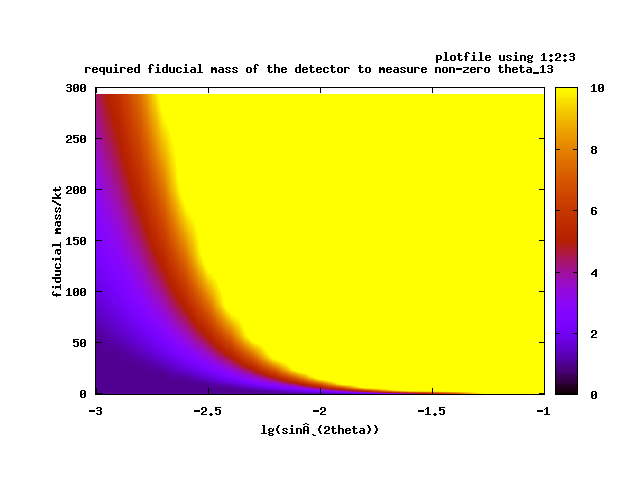}
\caption{\sf The $\theta_{13}$ reach ($3\sigma$) as a function of the detector at 2288 km baseline, vertical and horizontal orientation. The vertical direction costs a few kton.}
\label{reach_fid_VH}
\end{center}
\end{figure}

\begin{figure}[tbp]
\begin{center}
\includegraphics[angle=0, width=7cm]{theta23_15.png}
\includegraphics[angle=0, width=7cm]{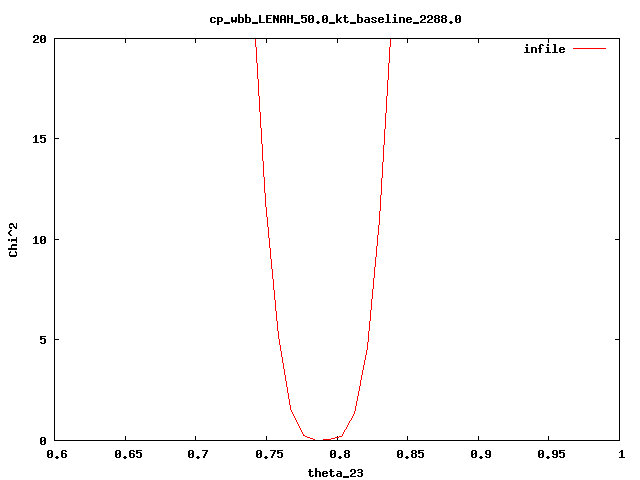}
\caption{\sf The accuracy of $\theta_{23}$ for 50 kton LENA in vertical and horizontal orientation. The difference is small but visible.}
\label{octant_VH}
\end{center}
\end{figure}

\begin{figure}[tphb]
\begin{center}
\includegraphics[angle=0, width=7cm]{wrong_sign_plot_LS.png}
\includegraphics[angle=0, width=7cm]{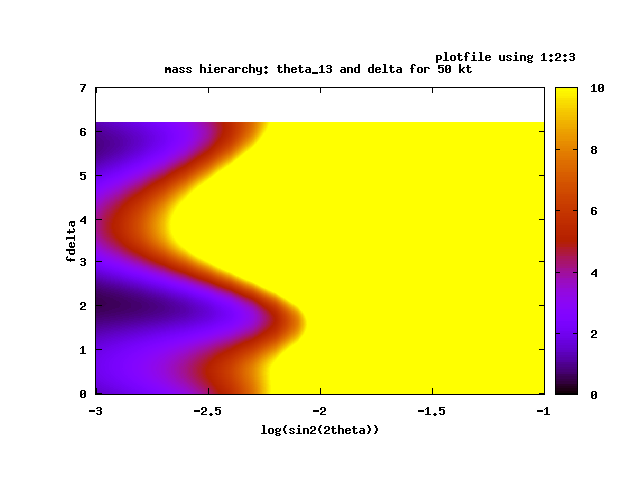}
\caption{\sf Comparison of the performance of the mass hierarchy for vertical and horizontal orientation: 2288 km.}
\label{mh_VH}
\end{center}
\end{figure}

\begin{figure}[htbp]
\begin{center}
\includegraphics[angle=0, width=7cm]{wrong_sign_fid_L0.png}
\includegraphics[angle=0, width=7cm]{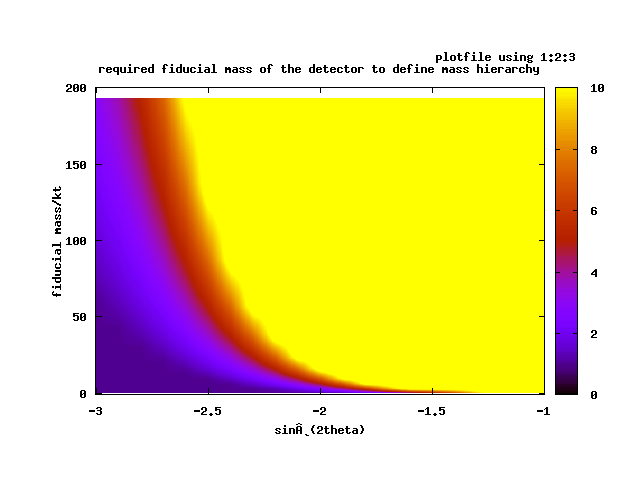}
\caption{\sf The capability to observe the mass hierarchy as a function of detector mass (scaling variable), for vertical and horizontal orientation (50 kt at 2288 km baseline, $\delta=\pi/2$).}
\label{mhori}
\end{center}
\end{figure}

\begin{figure}[tphb]
\begin{center}
\includegraphics[angle=0, width=7cm]{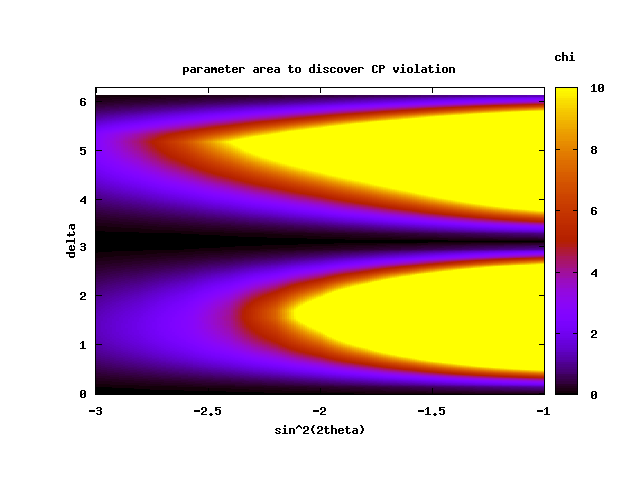}
\includegraphics[angle=0, width=7cm]{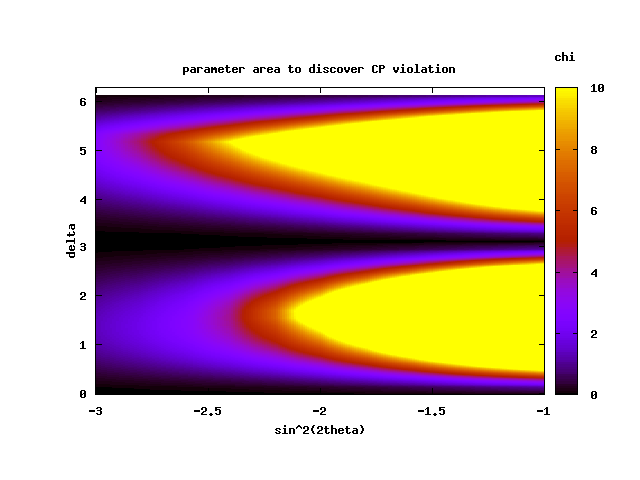}
\caption{\sf Comparison of the CP range for the vertical and the horizontal orientations, with 50 kton at 2288 km baseline. Differences are close to negligible.}
\label{deltasVH}
\end{center}
\end{figure}

\clearpage
\subsection{Significance of energy resolution}

\begin{figure}[hpb]
\begin{center}
\includegraphics[angle=0, width=7cm]{hierarhcy_eres_percent.png}
\includegraphics[angle=0, width=7cm]{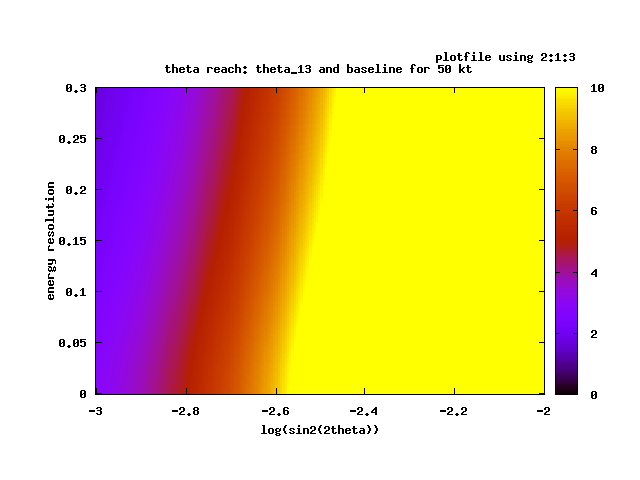}
\caption{\sf Comparison of the performance for measuring the mass hierarchy for different energy resolutions, relative given as per cent (left) and absolute in GeV(right).}
\label{mh_resos}
\end{center}
\end{figure}

\begin{figure}[hbp]
\begin{center}
\includegraphics[angle=0, width=7cm]{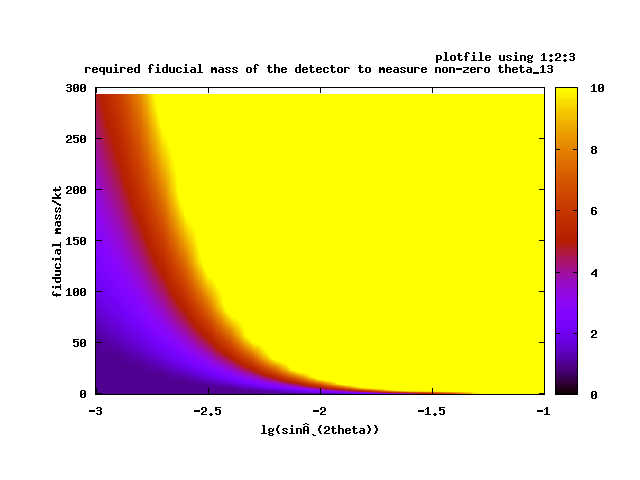}
\includegraphics[angle=0, width=7cm]{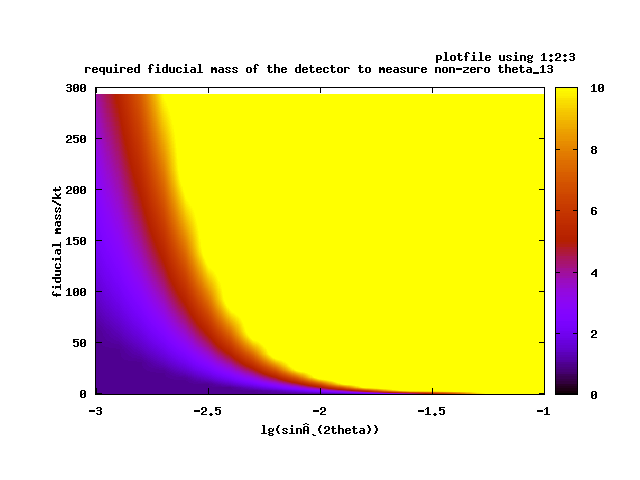}
\includegraphics[angle=0, width=7cm]{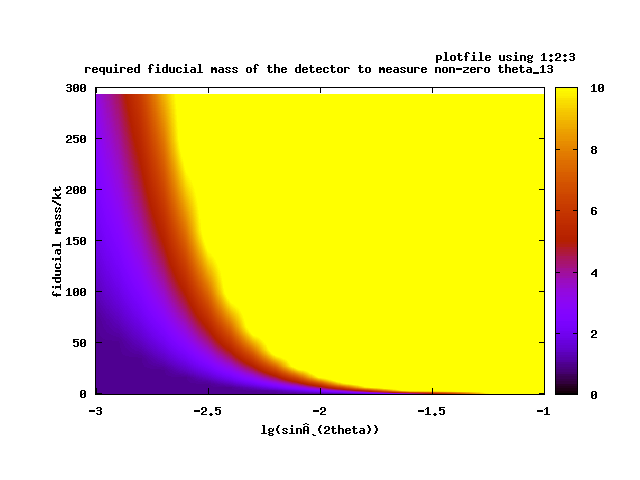}
\includegraphics[angle=0, width=7cm]{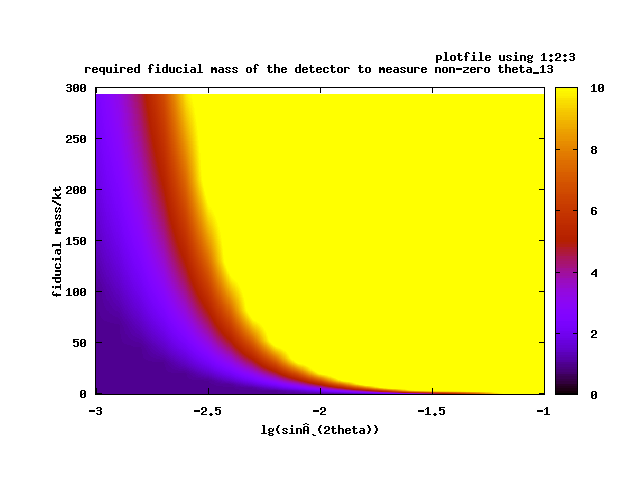}
\includegraphics[angle=0, width=7cm]{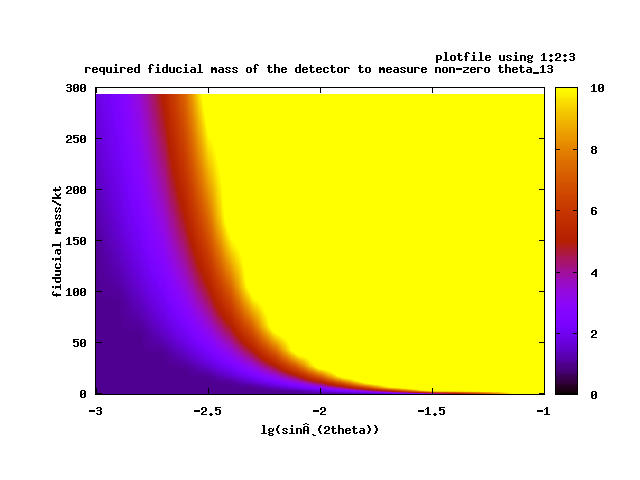}
\caption{\sf The $\theta_{13}$ reach ($3\sigma$) as a function of the fiducial mass of the detector at 2288 km baseline, for different energy resolutions of the detector (1 \%, 5\%, 10 \%, 20 \%, 30 \%). The differences are rather small for less than 10 \%, and hence a moderate resolution is sufficient.}
\label{reach_res}
\end{center}
\end{figure}

\begin{figure}[tbhp]
\begin{center}
\includegraphics[angle=0, width=7cm]{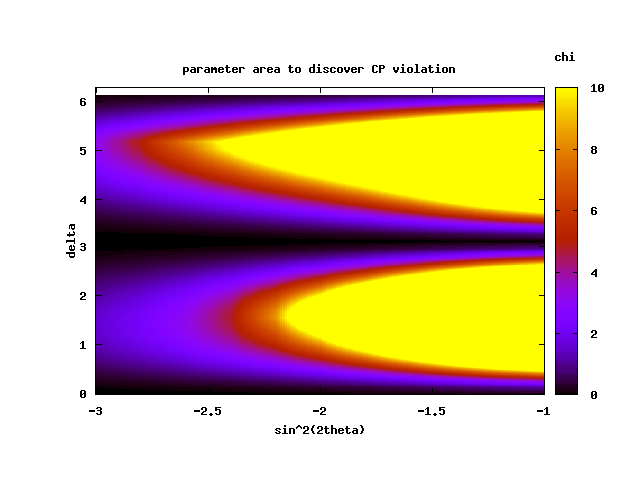}
\includegraphics[angle=0, width=7cm]{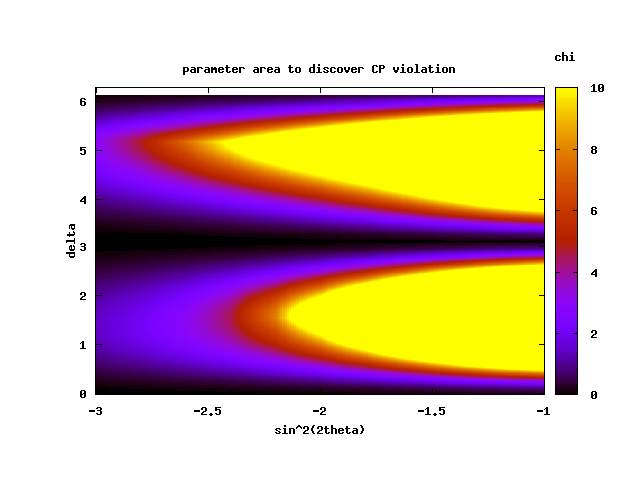}
\includegraphics[angle=0, width=7cm]{CPrange_rel_05.png}
\includegraphics[angle=0, width=7cm]{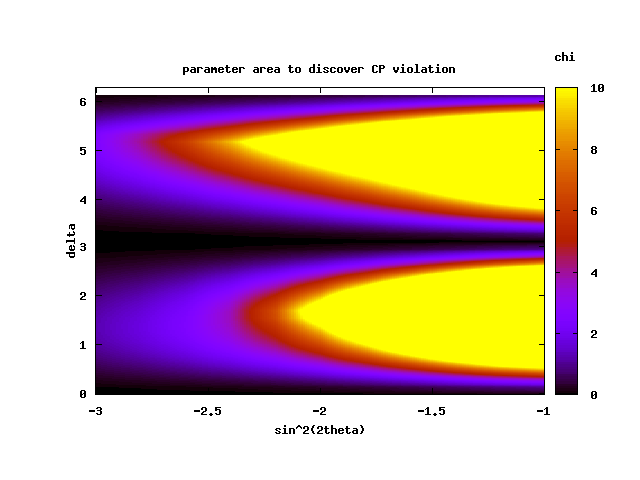}
\includegraphics[angle=0, width=7cm]{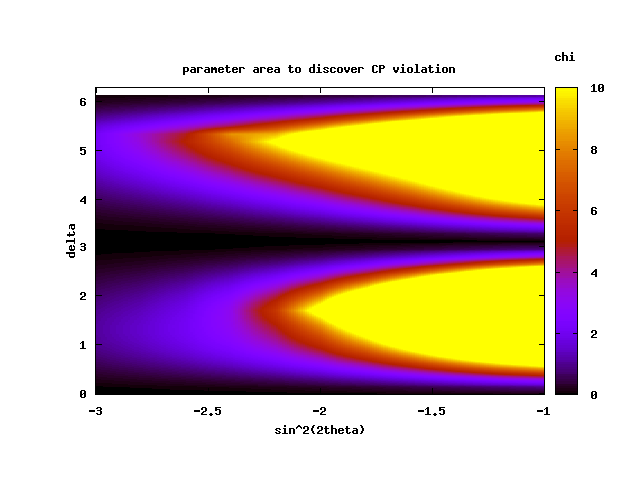}
\includegraphics[angle=0, width=7cm]{CPrange_rel_20.png}
\caption{\sf Comparison of the CP range with 1 \%, 3 \%, 5 \%, 7 \%, 10 \% and 20 \% relative energy resolutions. This shows that the results are not that sensitive to the energy resolution better than 5 \%, but worsening the resolution beyond 10 \% weakens the performance.}
\label{deltasdetq2}
\end{center}
\end{figure}

\clearpage
\subsection{Significance of the reduction of the wrong-sign muon background}

There may be some possibility to distinguish neutrinos from antineutrinos statistically. With lower energies this is more possible than higher. Here one case is with no distinguishion and the other for idealistic distinguishion, just for comparison. The difference seems negligible, and hence the antineutrino identification is not used in other studies.

\begin{figure}[htbp]
\begin{center}
\includegraphics[angle=0, width=7cm]{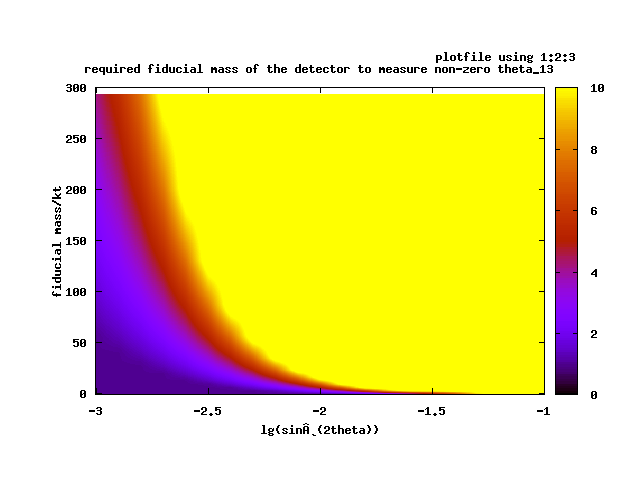}
\includegraphics[angle=0, width=7cm]{theta_reach_fid_L0.png}
\caption{\sf The $\theta_{13}$ reach ($3\sigma$) as a function of the detector at 2288 km baseline, for different antineutrino recognitions (perfect and zero). The differences are small.}
\label{reach_W}
\end{center}
\end{figure}

\begin{figure}[htpb]
\begin{center}
\includegraphics[angle=0, width=7cm]{CPrange0.png}
\includegraphics[angle=0, width=7cm]{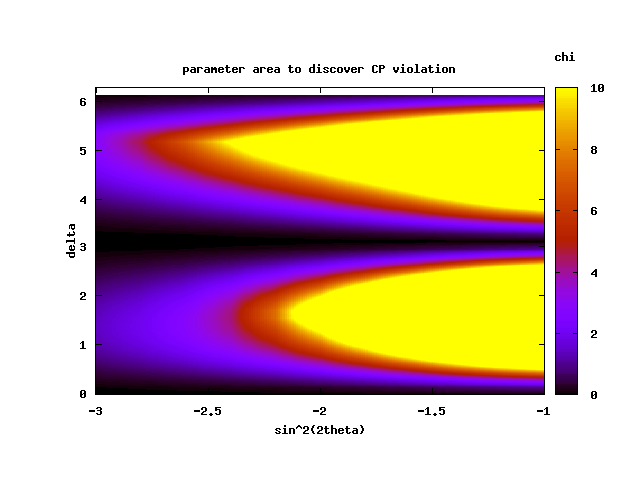}
\caption{\sf Comparison of the CP range for standard parameters and with full wrong-sign background reduction,
 all with 2288 km baseline. This shows that the results are not sensitive to the wrong sign muon background and the charge recognition is useless.}
\label{deltasdetWS}
\end{center}
\end{figure}

\clearpage
\subsection{Neutral current background}

\begin{figure}[tbhp]
\begin{center}
\includegraphics[angle=0, width=7cm]{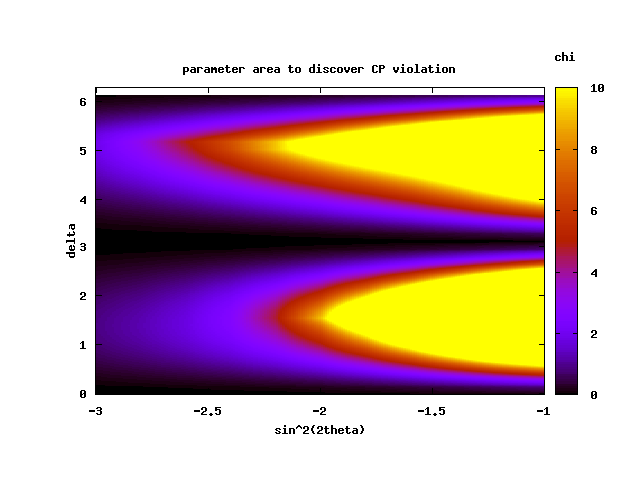}
\includegraphics[angle=0, width=7cm]{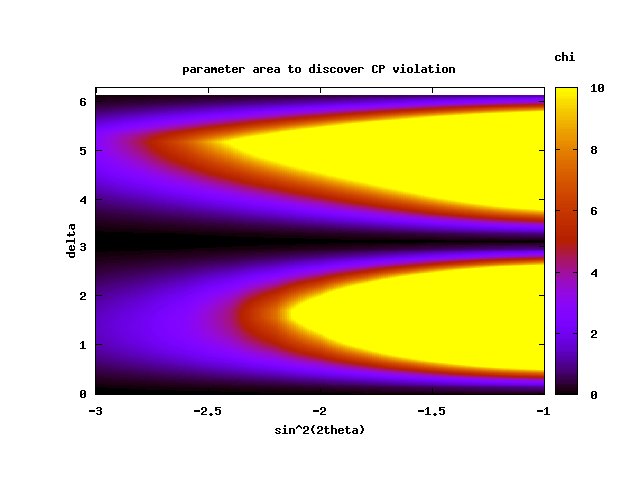}
\caption{\sf Comparison of the CP range assuming 10 times larger neutral current background (left) and no neutral current background at all (right). The latter is almost equal to the default case, so improving the neutral current event recognition provides no benefit, but a larger background decreases the capacity. }
\label{deltasdetNC}
\end{center}
\end{figure}

\begin{figure}[bhtp]
\begin{center}
\includegraphics[angle=0, width=7cm]{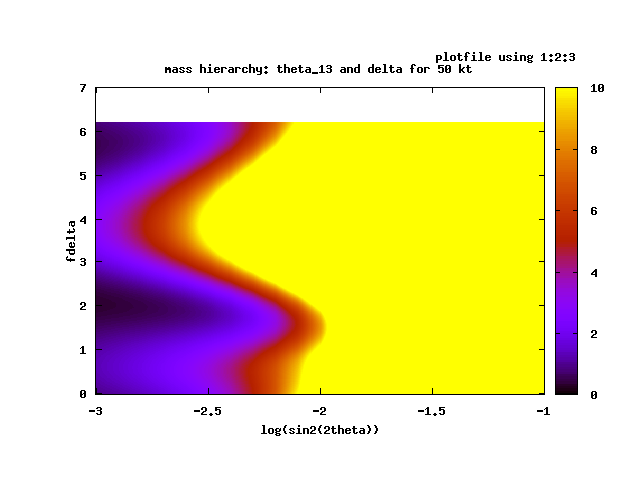}
\includegraphics[angle=0, width=7cm]{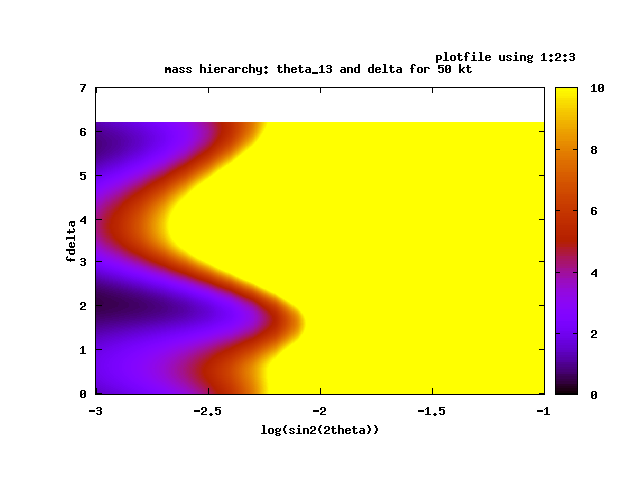}
\caption{\sf Mass hierarchy, for 10 times larger neutral current background (left) and no neutral current background (right) (50 kt at 2288 km baseline). Increasing the background weakens the performance visibly but not drastically}
\label{mhNC}
\end{center}
\end{figure}

\clearpage

\subsection{What if there were no beam background}

A few runs without beam and other background are shown. This is of course unphysical, but is done to show where the bottleneck at small mixing angles is. Evidently for small $\theta_{13}$ the performance is limited by the beam background. Other beams are required to extend the performance.

\begin{figure}[thbp]
\begin{center}
\includegraphics[angle=0, width=7cm]{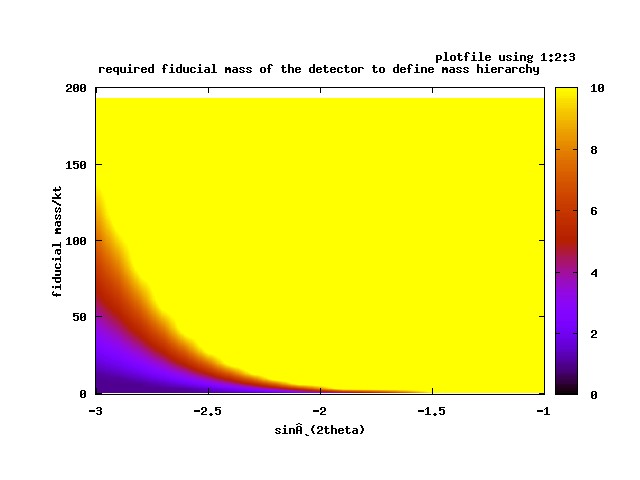}
\caption{\sf The capability to measure the mass hierarchy with a perfect detector with a hypothetical beam without background (at 2288 km baseline).}
\label{mhnobkg}
\end{center}
\end{figure}

\begin{figure}[bthp]
\begin{center}
\includegraphics[angle=0, width=7cm]{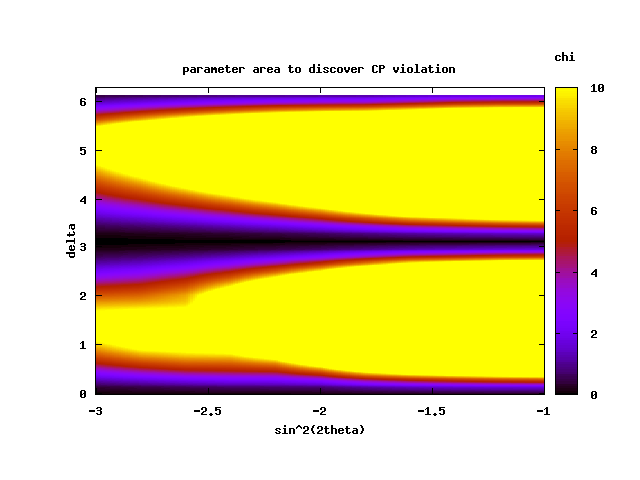}
\caption{\sf Test run for the CP range without any background.}
\label{deltasNB}
\end{center}
\end{figure}

\clearpage
\subsection{Comparisons with other baselines}

\begin{figure}[hbp]
\begin{center}
\includegraphics[angle=0, width=7cm]{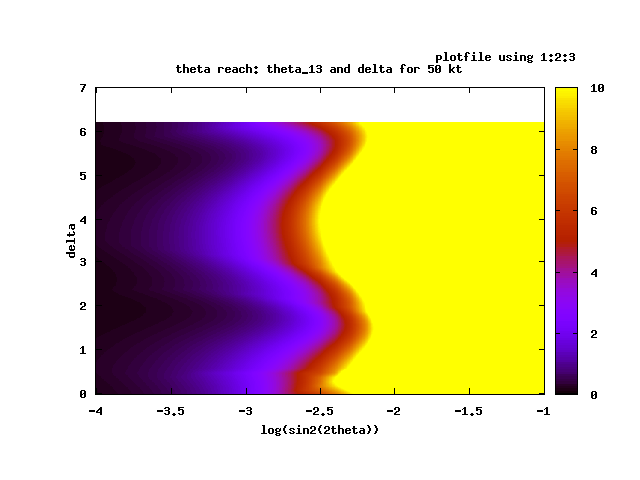}
\includegraphics[angle=0, width=7cm]{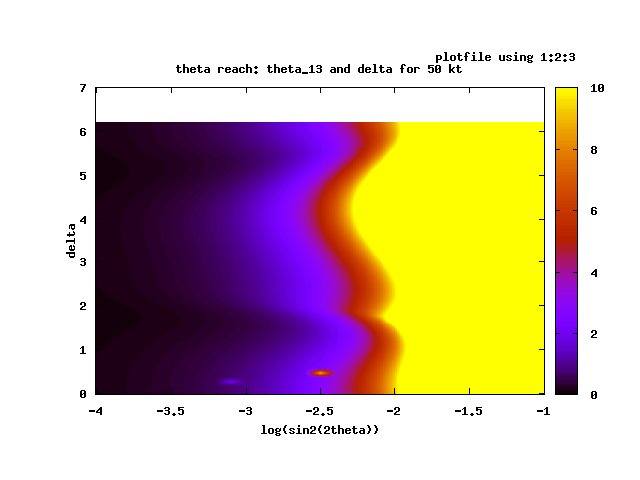}
\caption{\sf The $\theta_{13}$ reach ($3\sigma$) in $\delta$-plot for 50 kton LENA at baselines of 1544 km and 1050 km. While 1544 km baseline is very good, the shorter ones are less performant.}
\label{reach_baselines}
\end{center}
\end{figure}

\begin{figure}[hbtp]
\begin{center}
\includegraphics[angle=0, width=7cm]{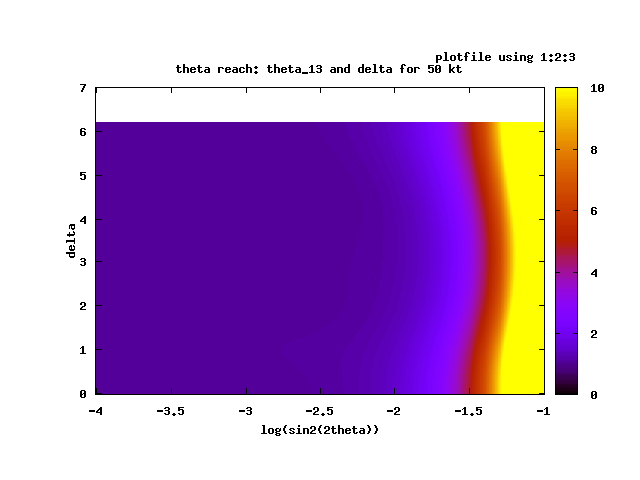}
\includegraphics[angle=0, width=7cm]{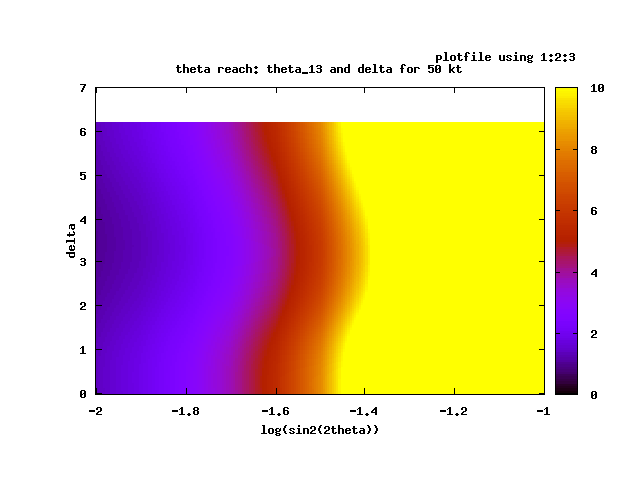}
\caption{\sf The range to define the octant of $\theta_{23}$ for 50 kton LENA at 2288 km and 1544 km baselines.}
\label{octant_baselines}
\end{center}
\end{figure}

\begin{figure}[tpb]
\begin{center}
\includegraphics[angle=0, width=7cm]{wrong_sign_plot_LS.png}
\includegraphics[angle=0, width=7cm]{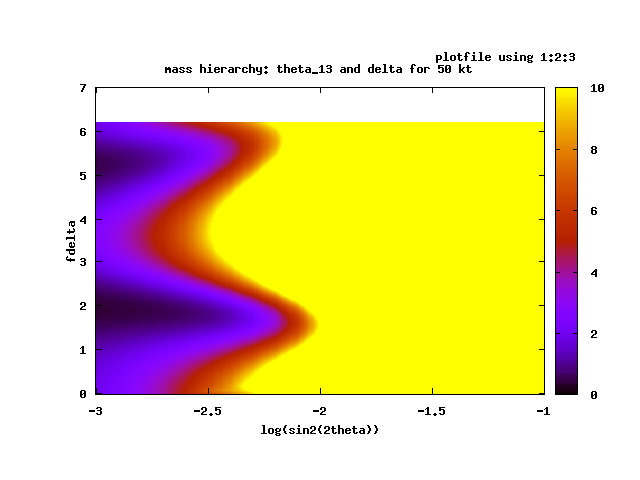}
\includegraphics[angle=0, width=7cm]{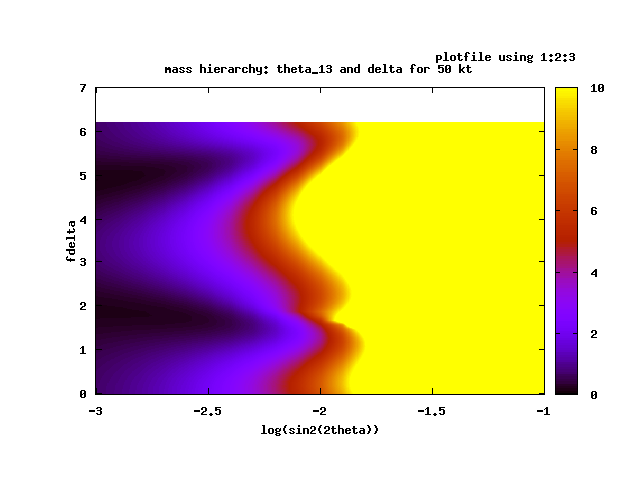}
\includegraphics[angle=0, width=7cm]{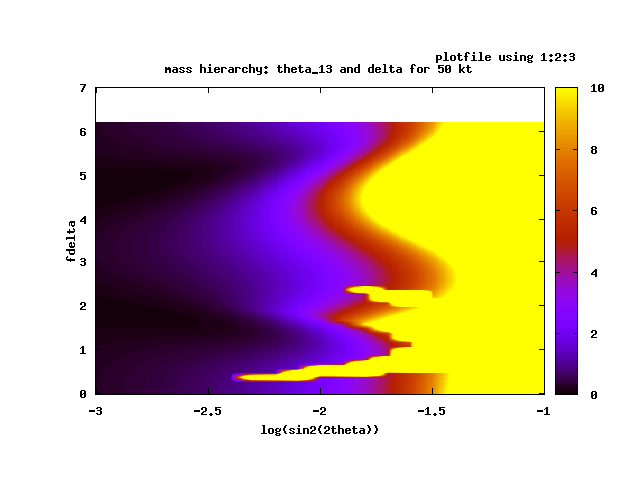}
\caption{\sf Comparison of the performance of mass hierarchy for different baselines: 2288 km, 1544 km, 950 km and 666 km.}
\label{WS_baselines}
\end{center}
\end{figure}

\begin{figure}[tbp]
\begin{center}
\includegraphics[angle=0, width=7cm]{wrong_sign_fid_L0.png}
\includegraphics[angle=0, width=7cm]{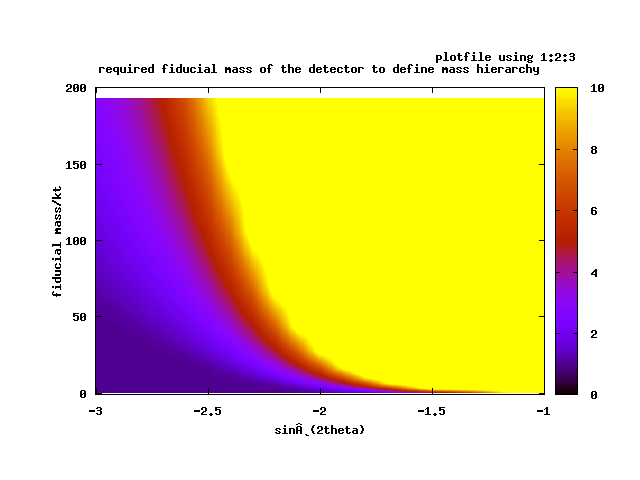}
\includegraphics[angle=0, width=7cm]{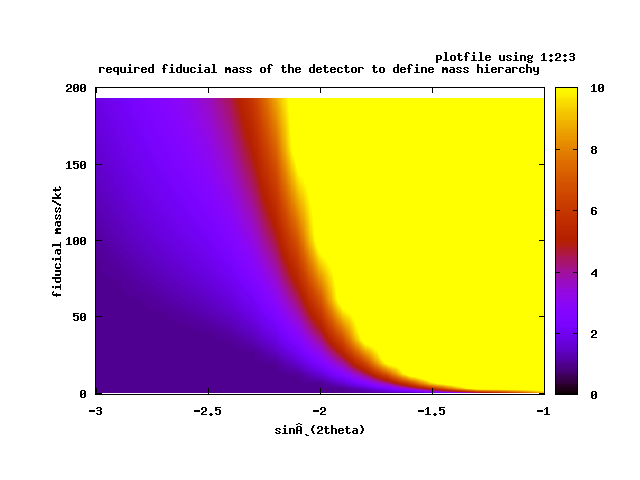}
\includegraphics[angle=0, width=7cm]{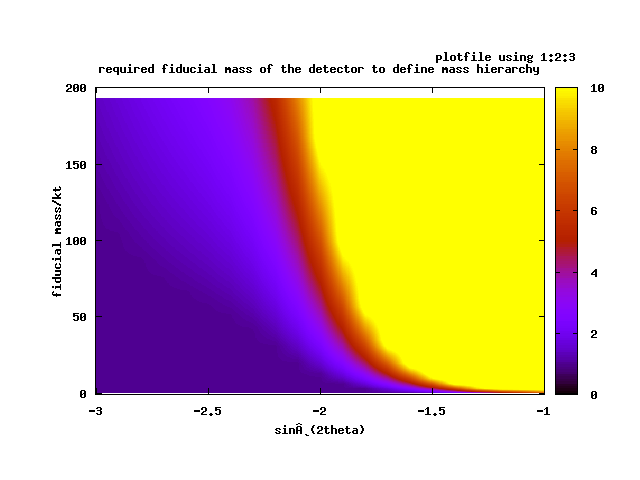}
\caption{\sf The capability to measure the mass hierarchy as a function of detector mass (scaling variable), for baselines of  2288 km, 1540 km, 1050 km and 950 km. Differences for longer baselines are minimal but baselines below 1000 km perform worse.}
\label{WS_fid_baselines}
\end{center}
\end{figure}

\begin{figure}[tpb]
\begin{center}
\includegraphics[angle=0, width=7cm]{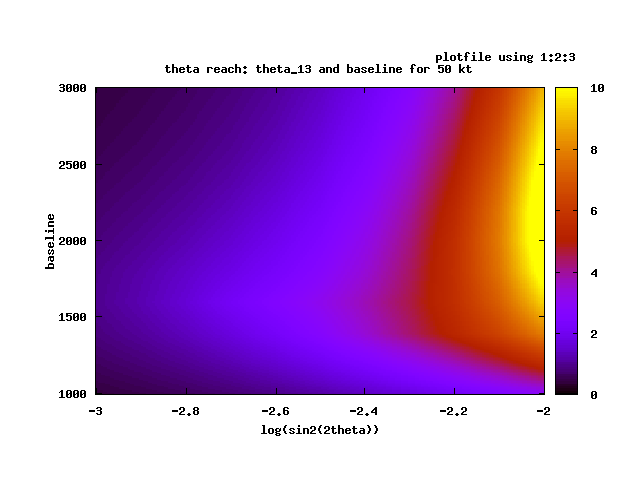}
\includegraphics[angle=0, width=7cm]{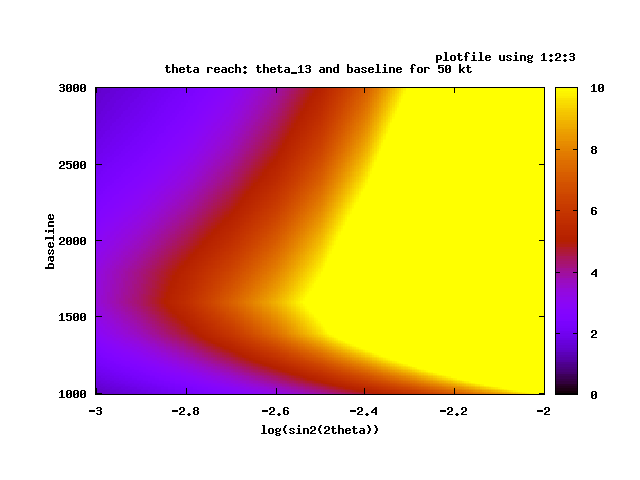}
\caption{\sf Comparison of the performance for the mass hierarchy for different baselines, for 25 kton and 100 kton}
\label{mhbaselines_25_100}
\end{center}
\end{figure}

\begin{figure}[tpb]
\begin{center}
\includegraphics[angle=0, width=7cm]{CPrange.png}
\includegraphics[angle=0, width=7cm]{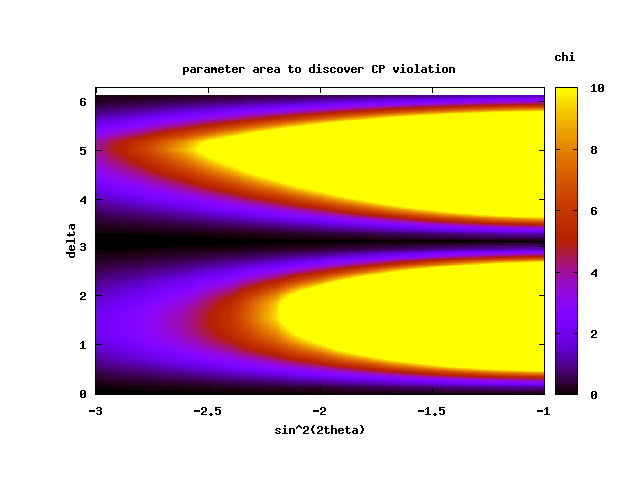}
\includegraphics[angle=0, width=7cm]{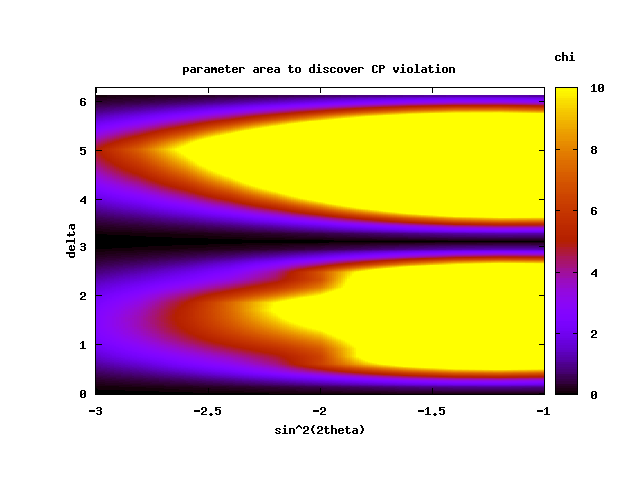}
\includegraphics[angle=0, width=7cm]{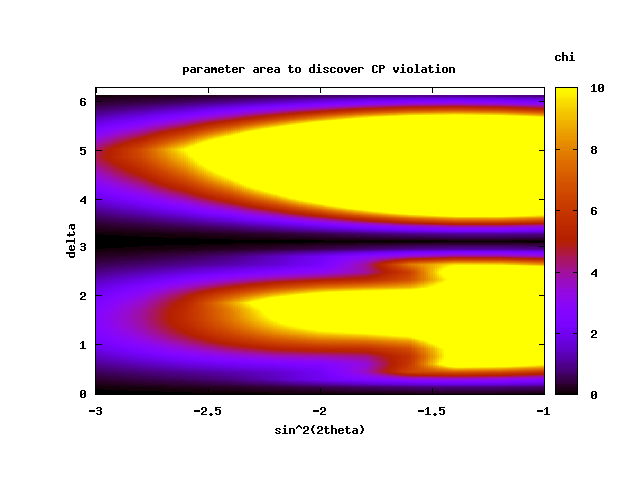}
\caption{\sf For comparison, the CP range with baselines 2288 km, 1544 km, 950 km and 666 km. For the last the performance is worse due to weaker sensitivity to mass hierarchy. All with 1\% uncertainty in density profile.}
\label{deltas_baselines}
\end{center}
\end{figure}

\end{document}